\documentclass[preprint,journal]{vgtc}
\usepackage[mathlines,switch]{lineno}

\ifpdf
  \pdfoutput=1\relax                   
  \usepackage{graphicx}                
  \DeclareGraphicsExtensions{.pdf,.png,.jpg,.jpeg} 
\else
  \usepackage{graphicx}                
  \DeclareGraphicsExtensions{.eps}     
\fi%

\graphicspath{{figures/}{pictures/}{images/}{./}} 

\usepackage{microtype}                 
\PassOptionsToPackage{warn}{textcomp}  
\usepackage{textcomp}                  
\usepackage{mathptmx}                  
\usepackage{times}                     
\usepackage{cite}                      
\usepackage{tabu}                      
\usepackage{booktabs}       
\usepackage[table, svgnames, dvipsnames]{xcolor}
\usepackage{scalerel}
\usepackage{changes}
\usepackage{mathtools}
\usepackage{enumerate} 

\usepackage{amsmath,amssymb,amsfonts}

\newcounter{lenumerate}
\renewenvironment{enumerate}{\begin{list} {\arabic{lenumerate}.}
{\usecounter{lenumerate} \setlength{\parsep}{0.1ex}
\setlength{\topsep}{0.6ex} \setlength{\partopsep}{0.1ex }
\setlength{\itemsep}{0.2ex} \setlength{\leftmargin}{2.4ex} }}
{\end{list}}

\renewenvironment{itemize}{\begin{list} {\labelitemi}
{\setlength{\parsep}{0.1ex} \setlength{\topsep}{0.6ex}
\setlength{\partopsep}{0.1ex } \setlength{\itemsep}{0.2ex}
\setlength{\leftmargin}{2ex} }} {\end{list}}

\onlineid{1524}

\vgtccategory{Research}

\title{Geo-Context Aware Study of Vision-Based Autonomous Driving Models and Spatial Video Data}

\author{Suphanut Jamonnak, Ye Zhao, Xinyi Huang, and Md Amiruzzaman}
\authorfooter{
\item S. Jamonnak (Email: sjamonna@kent.edu), Y. Zhao and X. Huang are with Kent State University, OH.
\item M. Amiruzzaman is with West Chester University, PA. 
}

\abstract{

Vision-based deep learning (DL) methods have made great progress in learning autonomous driving models from large-scale crowd-sourced video datasets. They are trained to predict instantaneous driving behaviors from video data captured by on-vehicle cameras.
In this paper, we develop a geo-context aware visualization system for the study of Autonomous Driving Model (ADM) predictions together with large-scale ADM video data. The visual study is seamlessly integrated with the geographical environment by combining DL model performance with geospatial visualization techniques. Model performance measures can be studied together with a set of geospatial attributes over map views. Users can also discover and compare prediction behaviors of multiple DL models in both city-wide and street-level analysis, together with road images and video contents. Therefore, the system provides a new visual exploration platform for DL model designers in autonomous driving. Use cases and domain expert evaluation show the utility and effectiveness of the visualization system.
} 

\keywords{Visualization System, Spatial Video, Autonomous Driving, Vision-based Deep Learning Models}

\teaser{
  \centering
  \includegraphics[width=\textwidth]{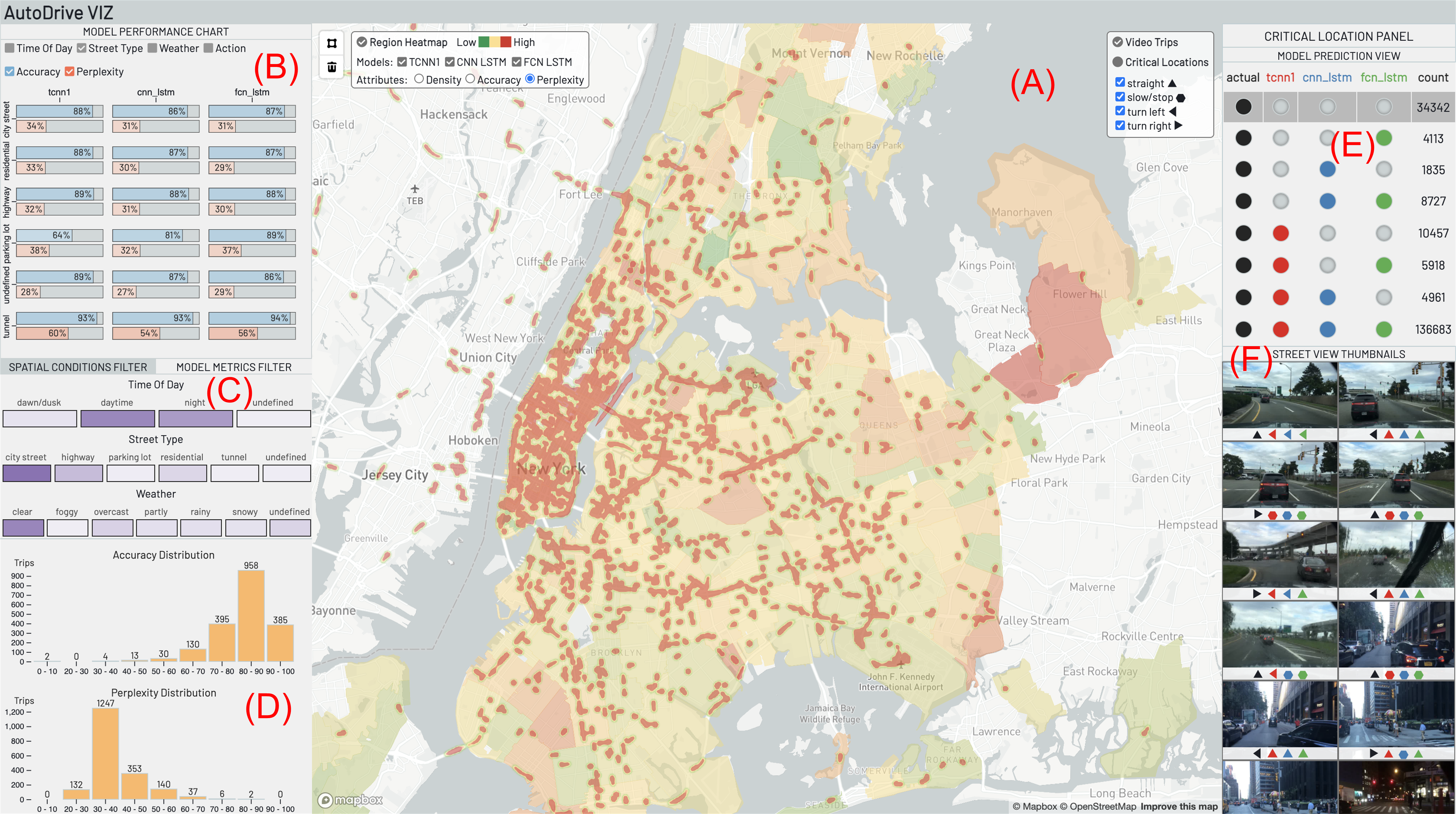}
    \caption{Visual interface for geo-context aware analysis of prediction data from autonomous driving models. (A) Interactive map view for deep learning model predictions and video data; (B) Autonomous driving model performance charts; (C) Trip filters based on model performance metrics or spatial conditions; (D) Trip distribution charts of selected video trip data; (E) Model prediction view for comparing multiple model predictions; (F) Street view thumbnails showing training video details at critical locations. } \vspace{-1pt} 
    \label{fig_interface} 
}

 \nocopyrightspace

\vgtcinsertpkg

\ieeedoi{10.1109/TVCG.20xx.xxxxxxx} 

\begin{document}

\firstsection{Introduction}
\maketitle

Deep learning (DL) in computer vision has achieved success in learning autonomous driving models (ADM) from large-scale crowd-sourced video datasets. The vision-based models are trained to predict instantaneous driving behaviors from spatial video data captured by on-vehicle cameras while moving vehicles traverse road networks in different types of built environments. The massive video datasets present diverse visual appearances, dynamic traffic situations, and meanwhile register realistic drivers' behaviors along the trajectories. The datasets are utilized for the development of DL models in autonomous driving. Domain researchers and practitioners of autonomous cars are continuously accumulating video data over multiple regions and cities. Many datasets and models are publicly shared to promote research progress.

These data analytics topics are of interest and importance to users including DL model designers and autonomous driving practitioners. Due to the large volume, complexity, and heterogeneity of the data, visual exploration techniques are demanded which can leverage the emerging DL research products in autonomous driving within geographical context. In this paper, we present a geo-context aware visualization system for the study of predictions made by vision-based ADMs together with large-scale video data. The scope of work is mainly for model performance comparison and analysis, and the target users are researchers of the related application fields.  

The system is built up based on an open repository of ADM videos including real driver actions and predictions from three different DL neural networks. A large number of video trips are processed and matched to geographical locations in a big city by their trajectory footprint locations. In a spatial database, the heterogeneous data of videos, images, DL model predictions are integrated with important attributes including streets, regions, weather, and time periods, and realtime queries are well supported using different conditions. Therefore, users are enabled to overview, search, and explore the performance of ADMs with geographical filters, and meanwhile, geographical locations with specified ADM performances can also be easily studied. A set of visualizations are designed for interactive study in both coarse and fine levels of geographical units. Video contents of traffic scenes are also visualized for model prediction analysis. Moreover, differences and similarities of three DL networks are classified and compared at locations, so that users can investigate specific model performances with geographic cues. 

In summary, the main contributions of this paper include:
\begin{itemize}
    \item We develop visual analytics (VA) techniques and a prototype that support domain users to study and compare multiple ADMs for their prediction performance in both city and street levels. 
    \item We present visual analysis functions that facilitate bidirectional data analysis, either from spatial conditions to ADM prediction performance or from model accuracy/perplexity to spatial attributes.
    \item We seamlessly integrate large-scale spatial video data and ADM prediction data within efficient data management and interactive geo-visualization interface. 
\end{itemize}
A few case studies were conducted with the new VA system in a big city area. Its utility and effectiveness were evaluated by domain experts. They agreed that the system fills the gap between the emerging large-scale ADM data and the analytical capability, which can be well utilized in their fields.

\section{Related Work}

\subsection{Vision Based Autonomous Driving DL Model}
Autonomous driving is an important developing technology that is useful for improving traffic, reducing emission, and transforming driving culture \cite{ADSurvey20}. Vision-based autonomous driving technology achieves fast and huge progress with DL models. Deep reinforcement learning\cite{Sallab_2017, jaritz2018endtoend, Perot2017} have been applied for motion control. The earliest ALVINN\cite{Pomerleau1988NIPS} uses a 3-layer shallow network with simulated road images for action prediction.
Later, a CNN is trained with road images and steering angles like Nvidia PilotNet\cite{bojarski2017explaining}.
More recent works for motion control are based on both CNN and RNN, where RNN is used for handling temporal sequences in video streams. FCN-LSTM\cite{xu2016end} uses a fully-convolutional network together with an LSTM to learn the visual and temporal information and make the decision on the action of motion.
Based on a CNN-LSTM architecture, Drive360\cite{hecker2018endtoend} further comprises a fully connected network to integrate information from multiple sensors for the prediction of the driving maneuvers.
The performance of deep learning models is measured quantitatively for statistical comparison. But there is a lack of interactive visual analytics tools to analyze the model performance with geographical attributes, which is the focus of this paper.

\subsection{Visual Analytics of DL Models}
Interactive visualization tools have been used for an in-depth understanding of how deep learning models work \cite{Hohman2019,Rauber2017}. Many VA tools (e.g., \cite{Wongsuphasawat2018,ConvNetJS,LSTMVis2018,RetainVis2019,RNNbow2019}) allow users to interact with the activation maps and network structure, and the prediction/classification results. These tools have the potential but are not yet applied to DL networks of ADM, where for instance, LSTM and CNN model visualizations may be integrated with street-view perception data. 

For vision-based DL models, computational approaches of deep learning explanation have been addressed through a variety of algorithms \cite{Samek2019}. A general taxonomy classified them into three main categories \cite{Grun2016}: input modification methods, deconvolutional methods, and input reconstruction methods. Deconvolutional Networks (DeconvNets) \cite{Simonyan2014,Zeiler2014}, Guided Back Propagation \cite{Springenberg2015}, Class Activation Mapping (CAM) \cite{Lin2014,Selvaraju2017,Zhou2016}, were the popular approaches. Recently, LRP has become an emerging focus from computer vision researchers (e.g., \cite{Arras2016,Sturm2016,Becker2018,Li2020}). Heatmaps were mostly used in these methods to visualize input pixels' relevance values to prediction results. VisLRPDesigner \cite{VisLRPDesigner21} provides a comprehensive visualization tool for LRP design and exploration. These methods and tools can be applied in the future to street-view-based ADMs to visually explain what perception features affect the driving action decisions.

Few VA systems are designed for autonomous driving models. VisualBackProp\cite{VisualBackProp2018} highlights network elements in Nvidia PilotNet\cite{bojarski2017explaining} that affect steering decision. VATLD\cite{VATLD2020} focuses on the understanding of the accuracy and robustness of traffic light detectors by disentangled representation learning and semantic adversarial learning. It allows users to obtain valuable insights to improve the CNN model performance. These tools do not study the prediction results of autonomous driving models and the spatial video data in the context of city-wide geography. Our system presents geographical visualizations together with model prediction metrics in a VA system, which enables users to conduct interactive studies among spatial, video, and model prediction data elements.

\subsection{Geospatial and Urban Data Visualization}

A variety of VA methods and tools have been developed to visually make sense of geo-spatial data  \cite{Andrienko2017}. They are developed for visualizing origin-destination movement data\cite{Andrienko2017,Natalia2013,Zhou2019}, vehicle trajectory data\cite{SemanticTraj}, space-time data\cite{Gennady2010,TopKube,Li2020}, flow maps \cite{1532150}, Flowstrates \cite{Boyandin:2011:FAV:2421953.2421999}, OD maps \cite{doi:10.1179/000870410X12658023467367}, and visual queries \cite{Ferriera13}. Vehicle trajectories are visually studied with various visual metaphors and interactions, such as GeoTime \cite{Kapler:2005:GTI}, SemanticTraj \cite{SemanticTraj}, TripVista \cite{Guo:2011:TTP}, FromDaDy \cite{FromDaDy12}, TrajGraph \cite{TrajGraph}, and more \cite{Qu:2011:Route, WangJam13,Wood:2007:IVE, adrienko2011spatial}. They have been widely used for urban and transportation data analytics, which however, have not been extended to leveraging the autonomous driving data.

On the other hand, street-view images have been used in landscaping, urban planning, transportation, and social studies. \cite{shen2017streetvizor,zhou2019social,ning2021sidewalk}. Recent development of DL technologies (e.g., Segnet \cite{badrinarayanan2017segnet} and PSPnet \cite{zhao2017pyramid}) make this process less expensive and faster, which can find objects and extract semantic categories from street-view images and videos.
A VA system \cite{GeoVisuals}, GeoVisuals, is developed to interactively manage, visualize, and analyze spatial video and geo-narratives using a set of visualization widgets and interaction functions. Moreover, VA systems for model diagnosis based on multi-modal sensors (camera, lidar, radar) have been developed and adopted in the AD industry (e.g. https://avs.auto/demo/). 

In this paper, we present a new VA system for the efficient study of vision-based ADMs inside a geospatial visualization platform. Our system integrates model prediction visualization with city-wide geographic environment.


\section{Vision Based Autonomous Driving Models}
Automated driving in urban scenes presents big interest and challenge in researchers and practitioners, leading to the advent of deep learning models together with the development of massive autonomous driving datasets. A survey reviewed the studies regarding present challenges, system architectures, emerging methodologies and core functions including localization, mapping, perception, planning, and human-machine interfaces \cite{ADSurvey20}. Vision-based DL models trained by spatial videos of traffic scenes have achieved impressive advances, which are studied in this paper. Next, we introduce the basics of  autonomous driving models (ADMs) and video datasets.

\begin{figure}[t] \centering {\makebox[\columnwidth]{\resizebox{\columnwidth}{!}{\includegraphics{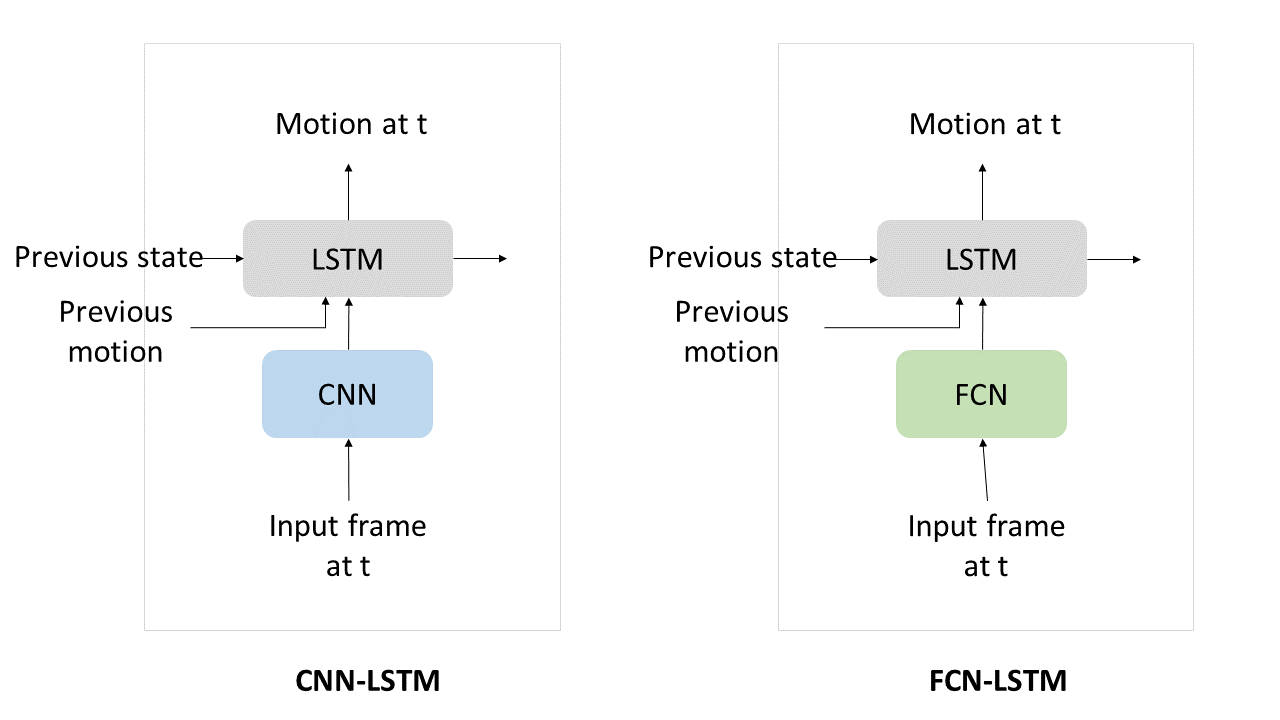}}}}\vspace{-5pt}
\caption{ An illustration of CNN-LSTM and FCN-LSTM networks.} \label{fig:models} \vspace{-15pt}
\end{figure}

\subsection{Discrete Action Driving Model}
Deep learning models used for autonomous driving action prediction usually take road view images from video cameras to predict possible future motion. Realtime image frames in a sequence of time steps are utilized to provide current and previous state signals, and the ADMs implement a driving action function as:
\begin{equation}
    G(i,a) \colon IMG \times ACT \rightarrow \mathbb{R}
\end{equation}
where $i \in IMG$ represents a set of images, and $a \in ACT$ represents a potential motion action. This function generates the probability of the action in real number. A typical set $ACT$ includes four discrete motion actions as:
\begin{equation}
   ACT = \left\{go ~straight, slow~ or~ stop, left~ turn, right~ turn \right\}.
\end{equation}
The set can also be extended to more actions (e.g., using slightly left turn, sharp left turn, and so on). These actions are usually qualified from the angular velocity and other car driving attributes. If the continuous velocity is used, it will lead to a continuous action driving model usually used for lane following problems. In this paper, we build the system based on the typical discrete action driving model.

\subsection{Model Training with Spatial Videos}

In order to design DL models for motion action prediction, large-scale spatial video datasets are needed for training and evaluation. Usually, crowd-sourced video datasets are collected from dashcam video cameras mounted on vehicles. The vehicles travel across roads and videotape various road conditions, while their GPS trajectories are stored. Meanwhile, real drivers' actions are recorded with sensors. It is demanded that these videos are generated in a large variety of geographical locations and settings, weathers, and time periods in a day, so that the DL models are well trained to face various traffic situations. Therefore, the DL model performance should also be studied based on these geographical attributes, which is one of our system goals.

There exist a group of open datasets, such as KITTI \cite{Geiger2013IJRR}, Cityscape \cite{Cordts2015Cvprw}, Comma.ai \cite{DBLP:journals/corr/SantanaH16}, Oxford \cite{RobotCarDatasetIJRR}, synthetic Princeton Torcs \cite{7410669} and GTA \cite{Richter_2016_ECCV}, which are publicly shared by ADM designers and researchers \cite{xu2016end}. In this paper, we utilize one of the largest datasets, the Berkeley DeepDrive Video (BDDV) dataset \cite{yu2020bdd100k}.

The BDDV dataset includes more than 10k hours of road videos in city, rural, and highway environments in multiple cities, multiple weather conditions (sunny, rain, snow, etc.), and both day and night times. It also comes with real driver actions and GPS trajectory data which can be mapped to locations at each time step. 

\subsection{DL Neural Networks}

A set of DL neural network architectures have been developed to make action predictions for the function $G$. Fig. \ref{fig:models} illustrates two networks. The input of image frame at current time $t$ is processed with either CNN (Convolutional Neural Network) or FCN (Fully Convolutional Network), and the output is combined in LSTM (Long short-term memory) network with previous states including the previous frames and actions in a few seconds prior to $t$. Then, the probabilities of motion actions in $ACT$ at $t$ is generated for prediction. The two models are named as CNN-LSTM and FCN-LSTM. In addition, a temporal CNN (TCNN) architecture can also be applied by adding an additional temporal convolutional module with a fixed time window \cite{xu2016end}. TCNN1 uses only a single image frame in the temporal window, so the two LSTM-fused models should have a better performance compared with TCNN1. We follow the methods in \cite{xu2016end} with CNN-LSTM, FCN-LSTM, and TCNN1 to show the visual study of multiple ADMs while other models can be further included. In particular, the three models are trained by video data from BDDV in New York City area in our prototype.

\subsection{Model Performance Metrics} \label{sec:metrics}
ADM prediction performance is usually measured by accuracy and perplexity. The accuracy is defined as
\begin{equation}
    accuracy = \frac {N_c}{N},
\end{equation}
where $N$ is the total number of predictions, and $N_c$ is the number of correct predictions when $argmax_{a} G(i,a)= a_{real}$. Here $a_{real}$ is the labeled action from the real driver. The number $N$ is usually counted for a whole test dataset to show the global accuracy of a model. In this paper, we realize that the accuracy can also be defined on a geographical region, so that the performance can be measured and studied within the geo-environment.

Model perplexity at time $t$ is defined over a sequence of $n$ predictions along the same driving path prior to $t$ as \cite{xu2016end}:
\begin{equation}
    perplexity = e^{- \frac{1}{n}  \sum_{k=1}^{n} \log G(i_k, a_k).  }
\end{equation}
Here, $i_k$ and $a_k$ are the $k$-th image frame and predicted action in the sequence. The perplexity is defined at each prediction and a low perplexity indicates more confidence in the prediction. An averaged perplexity of an ADM is often computed overall predictions in the whole test dataset. In this paper, a perplexity value is linked to the location where the prediction is made. Therefore, the average of perplexities at a geographical region or street can be computed interactively to indicate model performance with respect to geographical objects.

\section{Geo-Context Aware Data Processing}

\subsection{Video Trips and Data Extraction}
We download raw spatial video datasets from the public BDDV repository. In our prototype, we use the videos covering New York City (NYC) and its suburban area. The data is stored in the format of clips in the length of 40 seconds, with a frame rate at 30 fps and a high resolution at 720p. We call each clip as a \textbf{video trip}. Each video trip is associated with TripID, speed, GPS locations and timestamps, altitude, and other vehicle information. A total of 12,000 video trips with 130-hours of driving experience are used in our system. The total number of predictions is about 1.44 million. The video files are approximately 250GB in size. 

In addition to locations on the trace, each video trip contains spatial attributes including: (1) distinct times of day (day, night, dawn/dusk), (2) different street types (city streets, highways, residential, tunnel, gas station, parking lots), and (3) different weather types (clear, overcast, rainy, snowy, cloudy, foggy). The dataset contains approximately equal amounts of video trips in day-time and night-time. It also includes a large portion of weather conditions such as rain and snow. It is of great interest to study the ADMs with respect to these spatial attributes. For convenience in this paper, we call these attributes as \textbf{spatial conditions}, with an extended definition of ``spatial'' information.

\subsection{Data Processing and Transformation}
The extracted video trip data is processed in several steps: (1) generating TFRecords data; (2) making ADM predictions; and (3) computing ADM performance values.
\\
\textbf{Generate TFRecords:} To make efficient visual analysis, we transform raw data into TFRecords optimized for Tensorflow computing. The binary data format uses less disk space and less time to process. It is also essential to combine multiple data types for DL networks. For videos, ffmpeg library \cite{tomar2006converting} is used to extract frame images from video. Three images are extracted per 1 second. They are stored together with speed and timestamps in TFRecords. 
\\
\textbf{ADM prediction:} CNN-LSTM, FCN-LSTM, and TCNN1 networks are used to generate predictions of driver actions at each frame of these video trips, compatible with the recorded actual driver's action. The pre-trained DL networks \cite{xu2016end} are employed. For each frame, the neural network output includes a vector of prediction probability values for each action in \{1: go straight, 2: slow or stop, 3: turn left, 4: turn right\}. The action with the largest value is considered as the predicted action. For example for a frame image with the prediction values \{0.14, 0.35, 0.82, 0.46\}. The most probable action is ``turn left''. This action links to an accurate geo-location since the frame image has a recorded GPS location.
\\
\textbf{Computing model performance values:} The accuracy and perplexity values (see Sec. \ref{sec:metrics}) are finally computed for each frame image and link to the corresponding location as well. Here for a single image, the accuracy is either 1 or 0. For perplexity, it is computed from the previous seven predictions, that is, from seven previous image frames. These measures can be further aggregated to compute accuracy and perplexity for spatial units such as a trip and a spatial region.

\subsection{Map Matching and Spatial Database}
After data extraction and ADM computation, a set of locations along each video trip are linked to the predictions, performance values, actual actions, car speed, and spatial attributes. To enable geo-context aware visualization, these data items need to be matched with geographical objects such as streets, regions, and cities. Fast data query is needed to support interactive exploration. Therefore, we perform map-matching and then store and manage data in a specifically designed spatial database.
\\
\textbf{Map matching:} We first download road network geometry in NYC area from an open GIS data repository, OpenStreetMap (OSM) \cite{bennett2010openstreetmap}. We further retrieve zip code regions containing geometric boundaries. Each video trip is processed while each location with prediction results is mapped to a street segment and a zipcode region it resides in. In the process, some trips may go across different zipcode regions. Their GPS trajectories are cut into these regions respectively. 
\\
\textbf{Data management:}  A spatial database is devised to support data queries for interactive visualization. A NoSQL database is used specifically for the ADM and video data, instead of using a traditional relational database. The reason of choice is that the NoSQL database provides easy programming and efficient indexing for unstructured data including videos, images, and locations as a ``document''. It can also easily perform spatially based read or write operations on such a single data entity. A document encapsulates the location, timestamp, predictions, accuracy and perplexity, and various spatial attributes. It also includes links to the corresponding video clips and image frames, as well as the street segment and zipcode region it belongs to. This data structure can further incorporate other spatial and ADM information if needed. A video trip is further stored as a sequence of such locations. Moreover, the database also stores geo-structures of street networks and region geometries. In implementation, MongoDB is used for its popularity and easy to operate JSON files in data transfer.

The ADM prediction measures are further aggregated by averaging them over trips, streets, and zipcode regions. Furthermore, to support semantic query, we project the averaged accuracy and perplexity to multiple categories (bins). For example, a region with 64\% accuracy is categorized into the ``60-70'' percentage category. These categories facilitate easier visual exploration than raw numerical values. They can be used as \textbf{ADM performance conditions} for users to query data with different prediction performance ranges.
\\

\begin{table}[h] 
\caption{Querying by spatial and ADM performance conditions.}
\centering		{\makebox{\resizebox{\columnwidth}{!}{
\begin{tabular}{|c|c|}
\hline
\bf Query by            & \bf Spatial Conditions \\
\hline
Region          & Any polygon shape\\ \hline
Street  & Street Segment ID \\
\hline
Time of Day  & day, night, dawn/dusk \\ \hline
Street Type  & city street, highway, residential, \\
&tunnel, parking, gas station \\ \hline
Weather      & clear, overcast, snowy, rainy, cloudy, foggy \\\hline
\hline
\bf Query by            & \bf ADM Performance Conditions \\
\hline
Accuracy      & 1-10, 10-20, ... , 90-100 percentage\\\hline
Perplexity    & 1-10, 10-20, ... , 90-100 percentage \\
\hline
\end{tabular}
}}}\label{tbl:query}
\end{table}

\subsection{Query and Indexing} 
Table \ref{tbl:query} summarizes the queries used in our system based on spatial conditions and ADM performance conditions.
To enable fast data retrieval, spatial indexing is constructed in the database (MongoDB in default uses a B-tree subdivision of the space). Then geohash strings (e.g. \$geoWithin, \$geoIntersects) are used to quantify the locations to a cell in the tree. This scheme can quickly retrieve locations inside any queried region or streets by different conditions.

Furthermore, we create Boolean operational indexes for both accuracy and perplexity categories, and for spatial attributes. A variety of attributes thus can be combined in a data query using ``AND'' and ``OR'' operations. For example, the database can immediately respond to a query of video trips and locations where the ADM predictions happen at ``day OR night'' with a ``snow'' weather and the accuracy value ranges between ``30-40'' percentage. Such operations facilitate flexible data analysis tasks.

\subsection{System Implementations}
The datasets are processed on computers with Intel i7-8700K CPU and 16GB memory.  The predictions are conducted with TensorFlow, with GPU acceleration on either Nvidia GTX 1070 GPU with 8GB texture memory or Nvidia K80/T4 GPU with 12GB memory. The data processing time of each video trip including predictions of three ADMs cost on average about 5 minutes.

The visualization interface is implemented with Native JavaScript framework as client-side, bundling with several JavaScript libraries including Node.js, Mapbox-GL, and D3.js \cite{d3tvcg} libraries. The server-side script is implemented using Express.js and Mongoose.js JavaScript library in order to perform different query requests from MongoDB database. 

\begin{figure}[t]\centering
 \includegraphics[width=0.6\columnwidth]{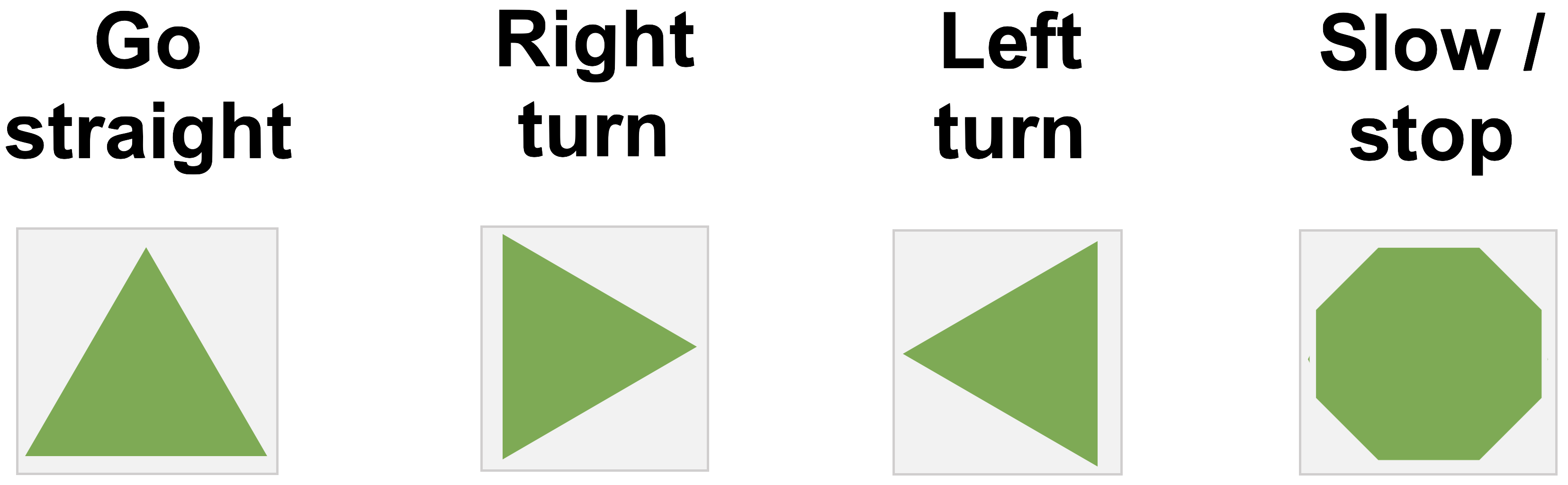}\vspace{-5pt}
    \caption{ Driving action icons. }\vspace{-15pt}
    \label{fig:icons}
\end{figure}

\begin{figure*}[t]\centering
 \includegraphics[width=0.98\textwidth]{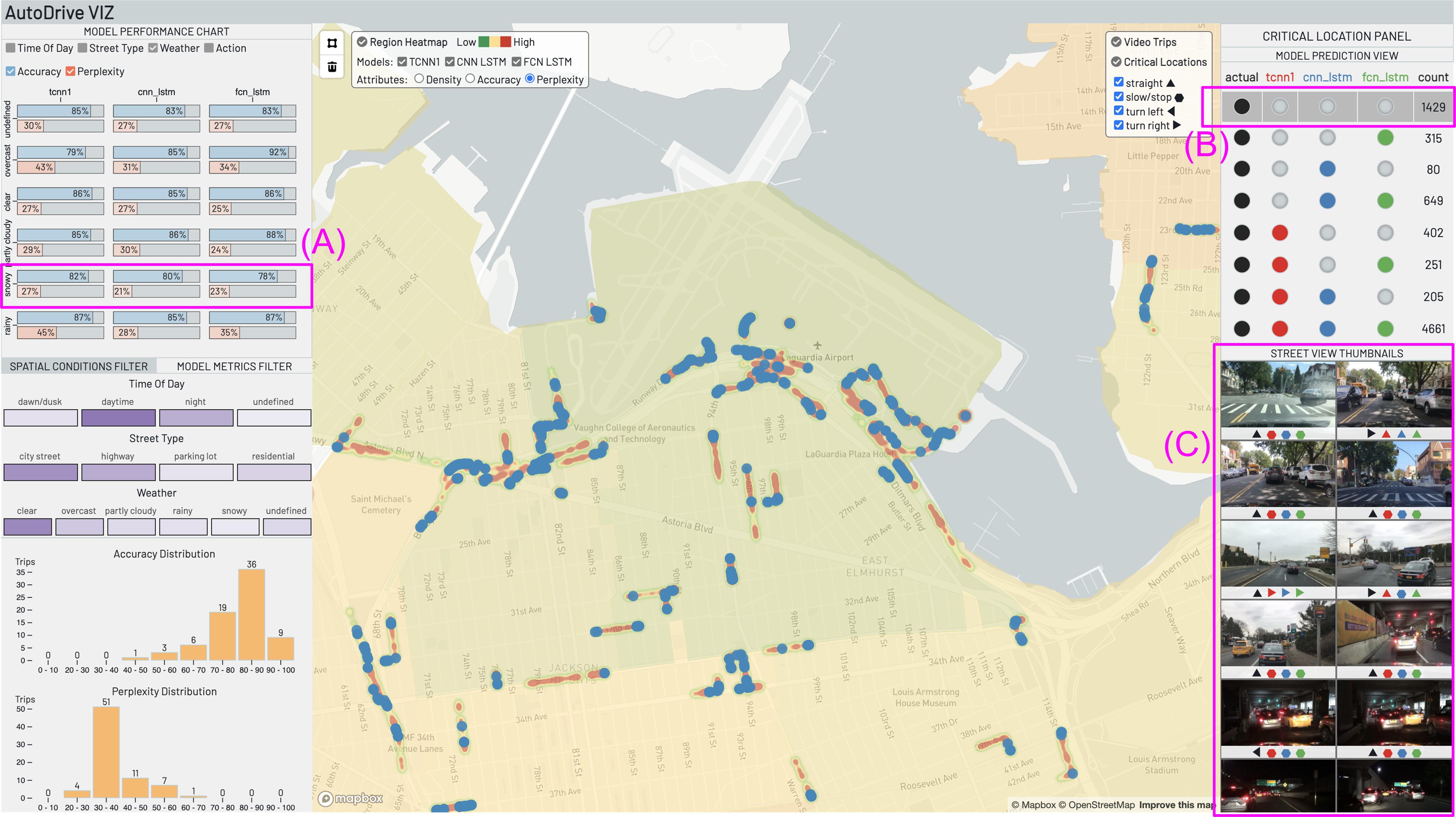}\vspace{-5pt}
    \caption{ Visual exploration of model performance with spatial conditions in a selected region. (A) Observe ADMs performance on snowy days; (B) Select wrong predictions of all three ADMs. Their locations are shown on the map as blue points; (C) Visualize corresponding video frames.  }\vspace{-15pt}
    \label{fig:case1_a}
\end{figure*}

\section{Design of The Visualization System}

\subsection{VA Task Characterization}

ADM researchers currently use statistical metrics such as accuracy and perplexity over benchmark datasets for model evaluation. They also compare predicted driver actions with actual driver actions by finding specific image frames in training datasets. However, their study has not been well supported by visual exploration, which can provide in-depth and interactive investigation connecting multiple models with locations and video contents. Therefore, our visual analytics system is designed for the following tasks:
\begin{itemize}
    \item \textbf{T1. Visually exploring model performance metrics:} Manage multiple models and a large set of training videos so that their performance metrics can be easily identified and understood. The exploration can be done with both large-scale analysis of global performance and fine-scale analysis of driver prediction behaviors. 
    
    \item \textbf{T2. Comparing different models:} Show and examine prediction results of various ADM implementations, so that users can easily compare these models and analyze their relations. 
    
    \item \textbf{T3. Analyzing models characteristics with visual contents:}
    Together with model performance study, distill and display related video contents and street-view environments, to provide important cues of critical situations that affect model prediction results.
    
    \item \textbf{T4. Studying models with spatial condition:} Integrate geo-context information into the above model analytical tasks by facilitating users with spatial conditions to perform ADM analysis, in the levels of streets and regions, and in the aspects of weather and time. Users can conduct efficient browsing, filtering, and queries, as well as a drill-down study of detailed information. 
    
    \item \textbf{T5. Studying locations with ADM prediction behaviors:} Discover geo-locations that have specified prediction results of interest. Users thus are able to link model behaviors to geo-spatial and/or environmental factors.
    
\end{itemize}
We realized that there exist more tasks in promoting wider use of ADM and its datasets, such as linking behaviors of neural network components with geospatial and street-view disparities, which can be further addressed in future work by VA community.

\begin{figure}[t]\centering
 \includegraphics[width=\columnwidth]{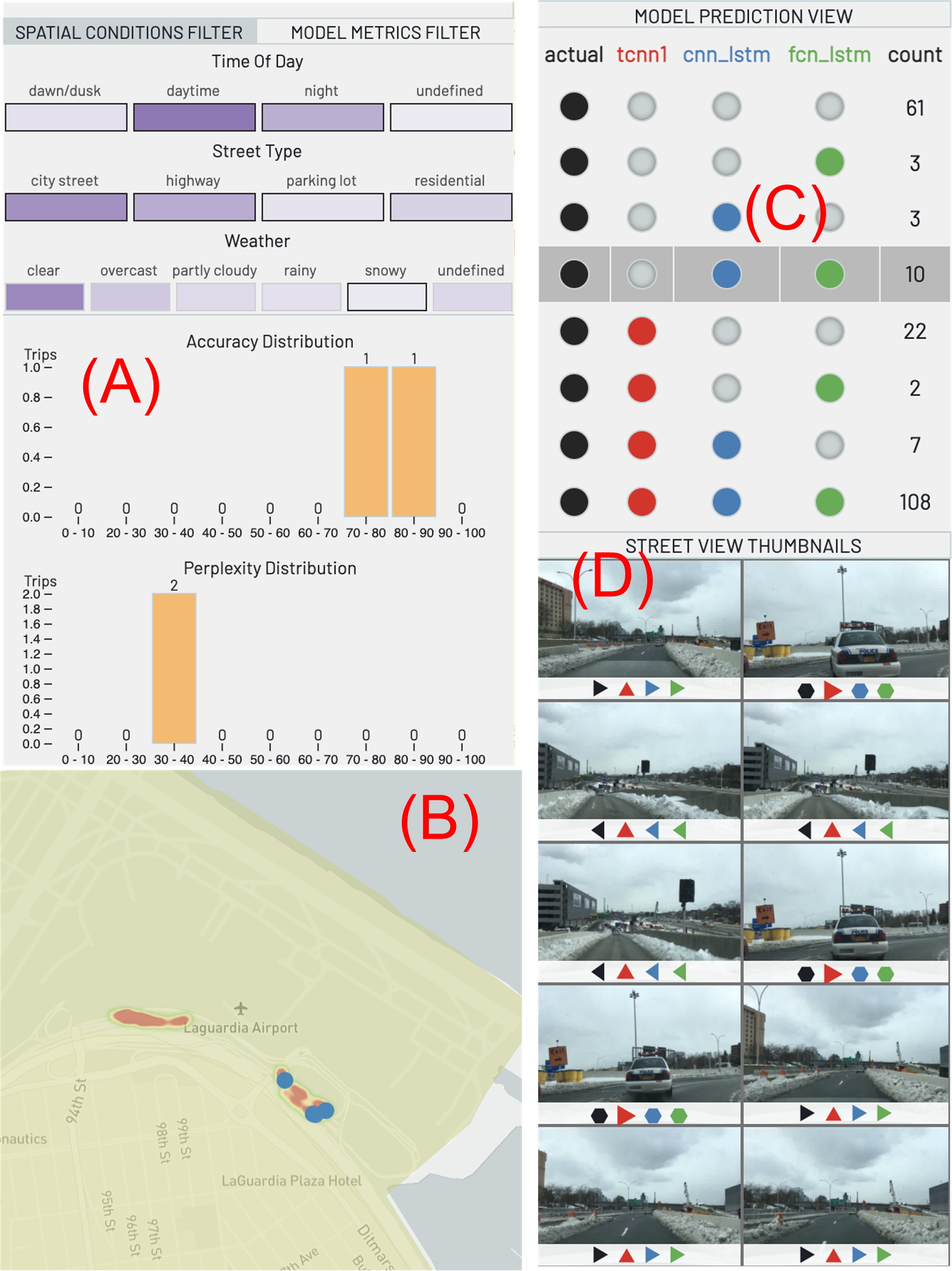}\vspace{-5pt}
    \caption{ Visual study of ADMs at snowy days in a local area. (A) View model performance distributions; (B) Visualize locations on map; (C) Study prediction features of ADMs; (D) Observe key video frames.  }\vspace{-15pt}
    \label{fig:case1_b}
\end{figure}

\subsection{Visualization Design}
With respect to the tasks, we design a visualization system which is illustrated in Fig. \ref{fig_interface}. The interface includes a set of coordinated views supporting interactive visual study:
\begin{itemize}   
\item {\bf Interactive Map View (Fig. \ref{fig_interface}A):} Geo-environment is visualized together with model performance and video data. Users can overview the whole city and can also select an arbitrarily sized region of interest. Here, three types of information visualizations are layered on the map: 
\begin{enumerate}
    \item Region heatmap: Zipcode regions are color-coded by one of the three ADM attributes including the density of locations where ADM predictions are made, ADM accuracy, and ADM perplexity. As shown in the legend on the top left corner of Fig. \ref{fig_interface}A, users can also select individual models for visualizing these attributes. 
    \item Video trips: Video trips are visualized in orange as trajectories. A kernel density estimation (KDE) algorithm is applied since many trips overlap in locations. The visualization shows their distribution in geo-space. 
    \item Critical locations: ADM locations are shown as blue points on the map, while users can select to show only those locations with specified driving actions, as shown in the legend on the top right corner of Fig. \ref{fig_interface}A.
\end{enumerate}
Users can make these layers visible or invisible on the map view. As shown in Fig. \ref{fig_interface}A, the region heatmap shows the average perplexity of all ADMs at zipcode regions in the NYC area. The available video trips in the area are also visible. The visualization shows that the benchmark video dataset has traversed most parts of Manhattan and many locations in other parts of NYC.

\item {\bf ADM Performance Chart (Fig. \ref{fig_interface}B):} 
Performance statistics of the three models, TCNN1, CNN-LSTM, and FCN-LSTM, are shown which are aggregated in realtime for any selected region on the map view. Users can select to check the accuracy or perplexity values of these models aggregated with respect to street type, weather, time of day. For example in Fig. \ref{fig_interface}B, the accuracy and perplexity are shown for different street types (highway, city street, etc.). In addition, users can also check the aggregated performance values according to actual driver actions.

\item {\bf Trip Filters (Fig. \ref{fig_interface}C) and Trip Distribution Charts (Fig. \ref{fig_interface}D):} 
Users can select video trips for bidirectional analysis. First, users can choose video trips based on spatial conditions in the \textbf{Spatial Conditions Filter} as shown in Fig. \ref{fig_interface}C. Then, the distribution charts of ADM accuracy and perplexity show the histograms of those trips satisfying the conditions (Fig. \ref{fig_interface}D). The trips are also shown on the map. Users can click on bars of the charts to further filter the trips. 

Second, users can also choose to use \textbf{Model Metrics Filter} to select trips based on the accuracy and/or perplexity values. For example, they can extract trips with an average accuracy smaller than 60\% to find questionable prediction results at spatial locations. With this filter, the Trip Distribution Charts show the histograms of the selected trips according to street type, weather, time of day. Users can interactively select specified bars on the charts as well. The filtered data will be updated on the map view.

\item {\bf Model Prediction Filter (Fig. \ref{fig_interface}E):} 
After trips of interest are selected, users can study and compare different model predictions results. In this view, four columns represent actual driver action, and predicted actions from the three models. If a model's prediction is the same as the actual action, the radio button is checked. Then, each row in this view represents a combination of predictions of different ADMs. The count of all predictions (i.e., locations and images frames) of each combination is shown. Buttons in the first column are all checked since makes it easy for comparison in rows with actual actions. For example, the first row in Fig. \ref{fig_interface}E indicates that there are 34,342 predictions in the selected trips which only FCN-LSTM gives a correct prediction. Users can also make critical locations of these predictions visible on the map to study them in detail.

\item {\bf Street View Thumbnails (Fig. \ref{fig_interface}F):} 
Visual contents of the critical locations are shown in the thumbnail view. Users can click to enlarge the image which is also linked to its position on the map. In the design of thumbnails, we realized that there are many images that are similar because predictions are made every 1/3 second from a video clip. Showing these sequential images cannot well utilize the visualization space. Therefore, we need to show different representative images at the top of the thumbnail view. Here, we implement an image comparison algorithm based on Structural Similarity Index Measure (SSIM), which automatically finds the most different images \cite{wang2003}. As shown in Fig. \ref{fig_interface}F, the thumbnail images show different locations and situations. Users can further see more images by sliding down this view. We limit the maximum number of video frames to be shown in this thumbnail view to 300. The choice is based on (1) users would not scroll down to check more than three hundred images; (2) the selected region in the spatial study usually do not have more than hundreds of different locations with different views; and (3) the interactive performance of the interface will not be affected.

Under each image, four \textbf{driving action icons} are designed to help users quickly identify the actual and predicted actions, which are shown in Fig. \ref{fig:icons}. The icons are colored as black (actual action), red (TCNN1 prediction), blue (CNN-LSTM prediction), and green (FCN-LSTM prediction), which are used throughout the system. 
\end{itemize}

In addition, users can conduct a drill-down study of videos and critical locations on video trips. A visualization interface of \textbf{Trip Study Interface} is shown in Fig. \ref{fig:case3}. It coordinates several views including:
\begin{itemize}
\item {\bf Regional Map View (Fig. \ref{fig:case3}A):} Users can click to select one trip or one location on the main interface, and then the regional map view automatically zooms into fine details around the selected object. A map inlet (Fig. \ref{fig:case3}B) shows this area's location in the global view. All the video trips are visualized as trajectories with a start point (green) and end point (yellow). The dataset does not have an excessive amount of trips in small areas, so it does not affect system performance by showing all trips.
Several trips may travel the same road. With this view, they can be clearly identified. An active trip can be selected which is highlighted and its driving direction is shown in black arrows. Users can also drag a marker to see details of the video and ADM prediction contents.

\item {\bf Trip List View (Fig. \ref{fig:case3}C):} This view lists all trips inside the zoom-in area. Each video trip's ID, prediction accuracy, and perplexity are shown for users to quickly check ADM performance. Users can click to select one trip in the list (or directly click on the map). Then this trip's video content is shown in Fig. \ref{fig:case3}D. Below the video, the actual speed and predictions of ADMs at the current frame are shown. Here in addition to the driving action icons, we also show the actual prediction probabilities. Users can further click them to show a popup view of the full prediction vectors (as shown in Fig. \ref{fig:case3}G).

\item {\bf Trip Timeline View (Fig. \ref{fig:case3}E):} Studying the variation of model performance over a trip from the trip start to end time is an important method. We design the timeline view where four rows indicate actual and three ADMs, respectively. At each row, the line chart shows the perplexity values at each corresponding location along this trip. And the dotted line separated by different action icons is used to indicate the prediction actions along the trip. For example, the bottom three rows have the predictions of stop/slow, then go straight, and then stop/slow again. But the predictions are not made at the same locations. A slider above allows users to drag along the trip to show accurate values of speed and prediction perplexities. It also changes the video contents in Fig. \ref{fig:case3}D and the marker location in Fig. \ref{fig:case3}A. 

\end{itemize}

A supplemental video is presented to show interactions on the interfaces and use cases. In this paper, the image and video contents have been modified by blurring when necessary to preserve privacy.

\begin{figure}[t]\centering
 \includegraphics[width=\columnwidth]{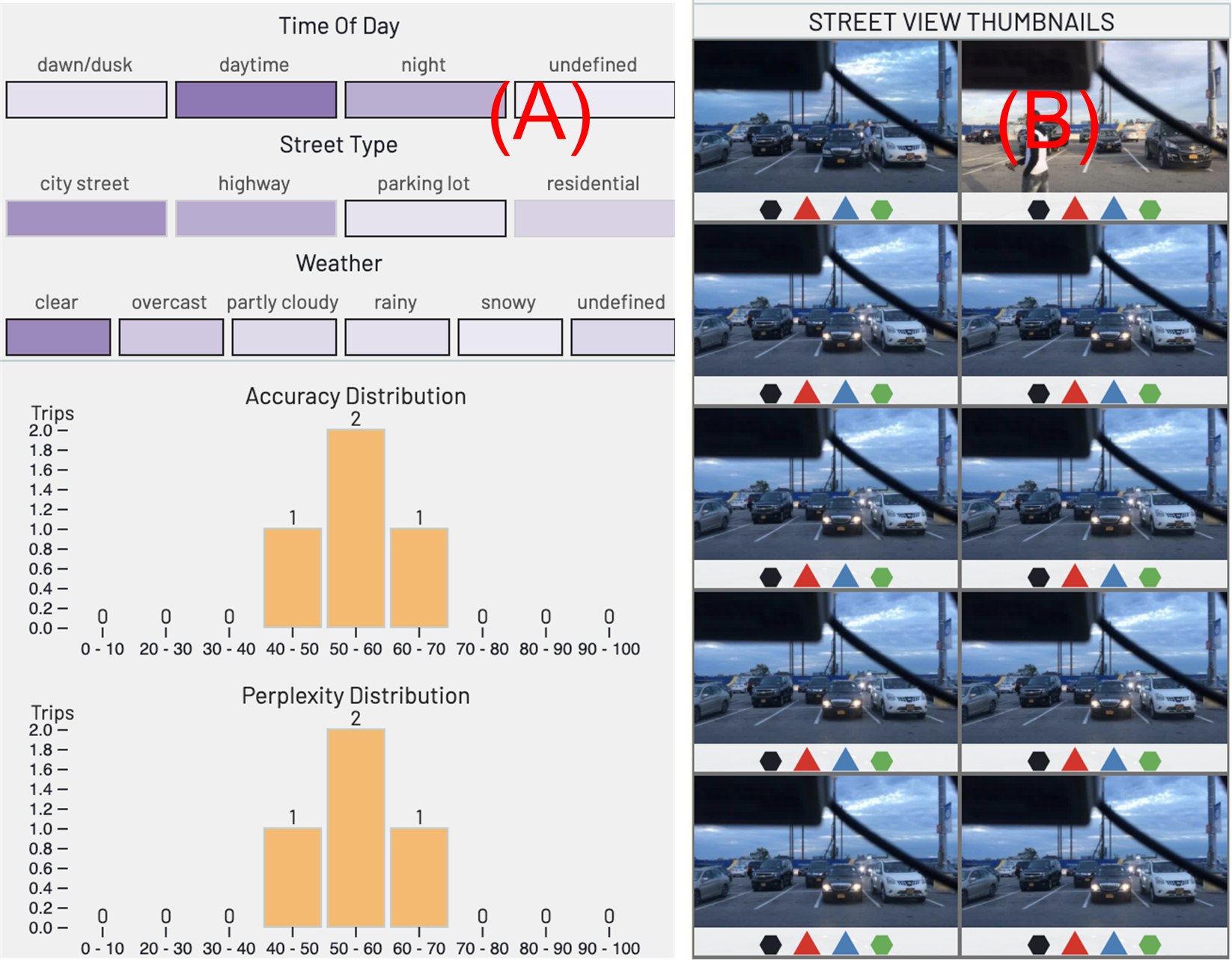}\vspace{-5pt}
    \caption{Visual study of ADMs at parking lots. (A) Filter data with paking lots; (B) Study related video details. }\vspace{-15pt}
    \label{fig:case1_c}
\end{figure}

\section{Use Cases}
A primary goal of autonomous driving researchers is to visually explore the performance of ADMs. Our system allows them to do this in different spatial regions. Next, we show examples of users performing visual exploration with our system.

\begin{figure*}[t]
\centering 
\centering 
\includegraphics[width=\textwidth]{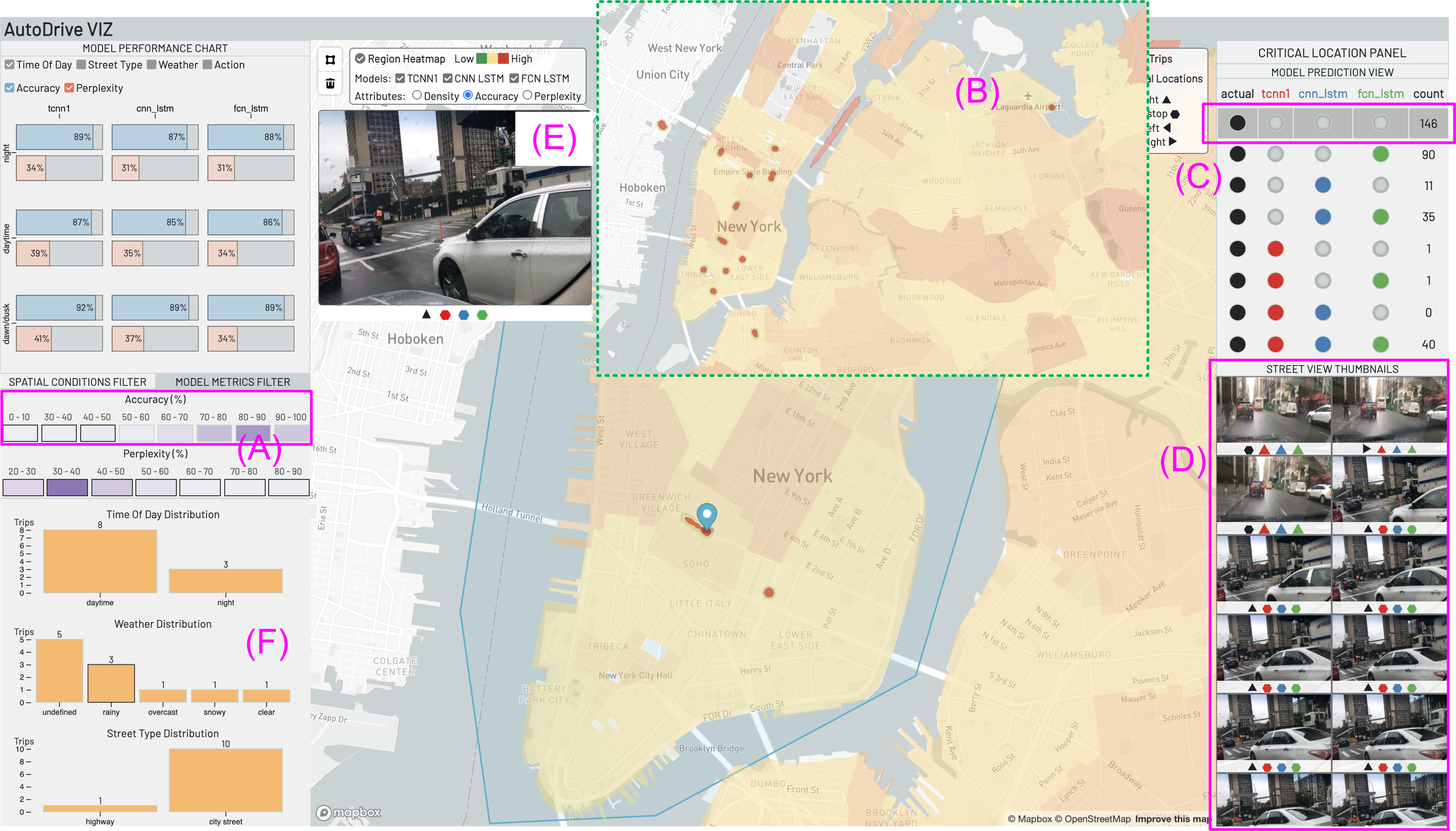}\vspace{-5pt}
\caption{Visual study of locations with model performance condition. (A) Select trips whose accuracy of three ADMs below 50\%; (B) Visualize spatial distribution of trips in the city; (C) Filter locations with wrong predictions; (D) Observe visual contents in thumbnails; (E) Enlarge view of selected thumbnail images; (F) Study data distributions with spatial attributes. }\vspace{-15pt}
\label{fig:case2}
\end{figure*}
\subsection{Case 1: Visual exploration with spatial conditions}

Fig. \ref{fig_interface} shows the global performance and street type information, together with video trips. First, users can observe spatial distribution of the data in the NYC area in Fig. \ref{fig_interface}A. It helps users understand that most videos are recorded in Manhattan and also cover major roads in Queens and Brooklyn. The training is related to the built-in environment in these areas. Thus, the trained models may not work well on other suburban or rural areas. In addition, the performance measures in the whole area are shown in Fig. \ref{fig_interface}B, where the accuracy and perplexity of the three networks are displayed with respect to street types. It can be found that they have similar values. But for parking lots, TCNN1 has relatively low accuracy. For tunnel locations, all the models have a high perplexity indicating low confidence. This knowledge can be used for model analysis and training improvement.

Users further observe the average accuracy and perplexity of zipcode regions in the area by hiding the trips on the map. For example, users draw a polygon around one region with low perplexity and zoom in to study it. It can be seen on the map that this area is around LGA airport.

As shown in Fig. 
\ref{fig:case1_a}, in this area, most of the video trips are in the highways and roads around the airport, while a few of them ar in the residential area. In Fig. \ref{fig:case1_a}B, users choose to show critical locations where all three ADMs fail in prediction, which appear as blue points on the map. By studying Fig. \ref{fig:case1_a}C in the street view thumbnails, users can click each location on the map or on the thumbnails to study individual situations. 

Users then study weather-based ADM performance, they find snowy day links to a low accuracy (Fig. \ref{fig:case1_a}A). Then a snowy day in the spatial conditions filter is selected as shown in Fig. \ref{fig:case1_b}A, finding two related video trips whose accuracy and perplexity distributions are displayed. The two trips are shown on the map (Fig. \ref{fig:case1_b}B). Users study on the model prediction view where one row is selected (Fig. \ref{fig:case1_b}C). This row reflects ten predictions where CNN-LSTM and FCN-LSTM give correct predictions and TCNN1 fails. The ten prediction locations are shown on the map and the street view thumbnails (Fig. \ref{fig:case1_b}D). It can be seen that snow on roadsides may affect the performance of TCNN1 while the traffic is very light.

Users can also conduct a study of street types. As shown in Fig. \ref{fig:case1_c}A, the video trips at parking lots (based on Fig. \ref{fig:case1_a}) are selected and they have low accuracy and high perplexity values. Users observe their locations which are in the outdoor airport parking lot. In Fig. \ref{fig:case1_c}B, a set of thumbnail images are shown. These images are part of the locations where only FCN-LSTM gives correct predictions and other ADMs fail. In these scenes, the actual driver's actions are all stop while TCNN1 and CNN-LSTM give predictions of go forward. It may be due to the lack of training for stop situations, and also shows the advantage of FCN-LSTM, which the original designer claimed as the best model \cite{xu2016end}.

\subsection{Case 2: Visual study of locations with model accuracy condition}\label{case:case2}

In addition to study ADMs with spatial conditions, it is also of interest to perform visual analytics based on performance metrics. As shown in Fig.\ref{fig:case2}, users can study locations related to low performance (accuracy $< 50\%$) of ADM prediction. In Fig.\ref{fig:case2}A, users select accuracy percentages lower than 50. In the overview of Fig.\ref{fig:case2}B, it is interesting to find that most of such trips are distributed in the lower Manhattan area. This observation matches the common opinion that it may be hard for autonomous vehicles to operate in complex built-in environments such as lower Manhattan. This area has a complex transportation network, situations and rules, high-rise buildings, and many pedestrians.  

After zooming in to this area, users select the top row in the Model Performance View (Fig.\ref{fig:case2}C), where all three ADMs give wrong predictions at 146 locations shown on the map (It can also be seen that the bottom row indicates only 40 locations have correct predictions). Users check the critical locations in the top row by studying their visual contents in Fig.\ref{fig:case2}D. They can click to enlarge them for details (such as in Fig.\ref{fig:case2}E). It can be realized that these wrong predictions happen in busy city streets. In Fig.\ref{fig:case2}E, one situation happens when stopped traffic just starts to move at an intersection. Here the actual driver's action is go ahead, and the ADMs predict slow and stop. This indicates the anticipation by the real driver that there is no need to slow based on prior experience. On the contrary, the ADMs rely on vision information only which may not easily give anticipated predictions in this situation.

Moreover, in Fig.\ref{fig:case2}F, users find these error predictions mostly happen at ``daytime'' in comparison to ``night'', which is possibly due to the rush hours in this busy city. And they happen mostly in ``undefined''(data missed) and ``rainy'' days and on ``city streets''. Such information may provide guidance for collecting future training data to improve model performance. 

\begin{figure*}[t]
\centering 
\centering 
\includegraphics[width=\textwidth]{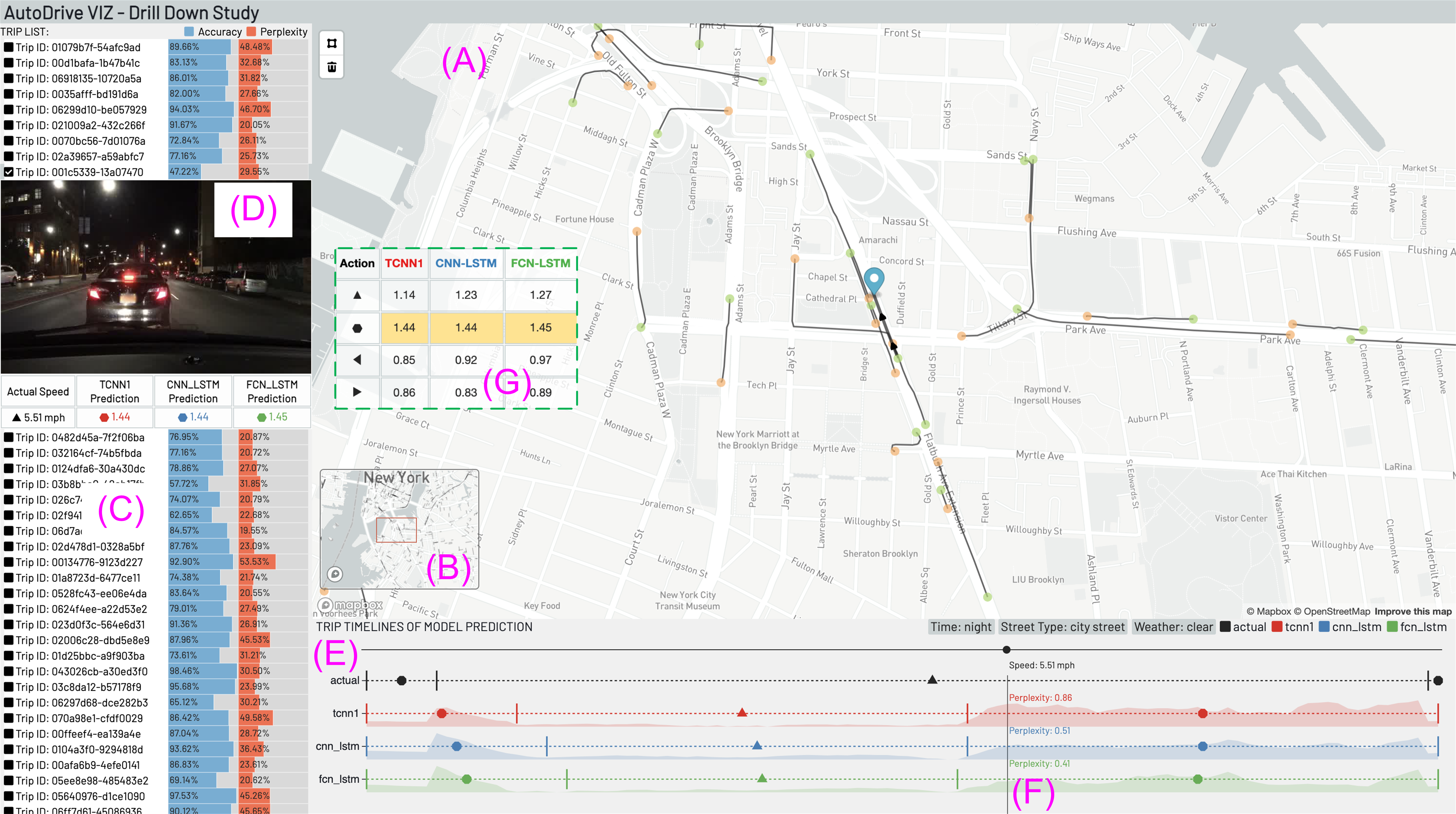}\vspace{-5pt}
\caption{Drill-down visual study of video trips. (A) Regional map view; (b) Global map inlet; (C) Video trip list view: (D) Corresponding video of a selected trip; (E) Trip timelines view of model prediction details; (F) Slider on the timelines to show detail performance values. (G) Pop-up view of prediction vectors. }\vspace{-15pt}
\label{fig:case3}
\end{figure*}

\subsection{Case 3: Drill-down visual study of video trips}

Fig. \ref{fig:case3} shows the drill-down visual exploration of video trips following the study in Fig. \ref{fig:case2}. Here, users select one trip on the map view (Fig. \ref{fig:case3}A) which goes from south to north along a major city street. The weather is clear and the time is night, which are shown at the bottom of the map. The video of this trip played in Fig. \ref{fig:case3}D. Users also observe the details in Fig. \ref{fig:case3}F to overview the ADM predictions of this trip. In the timeline view, users realize that there exists a sudden increase of perplexity values for all three ADMs. They drag the slider to these locations. The video is also played to the corresponding locations where a car ahead is braking. The three predictions give conflicting results to the actual action. While the real driver's action is go straight, the ADMs predict slow and stop. Here, the vehicle speed is low (5.51 mph) and so the real driver feels it is OK to go ahead. But the ADMs suggest preventive driving to slow/stop the car for safety. Users further click to see the details of prediction vectors, as shown in Fig. \ref{fig:case3}G. They find the prediction scores for ``stop/slow''(1.44 or 1.45) are just slightly higher than ``go straight'' (1.14 - 1.27), which partly explains the high perplexity. In such cases, maybe the models can be refined to include preventive driving actions through better training. 


\section{Expert Evaluation}

\subsection{Domain Experts and Procedure} 
We interviewed several domain experts to evaluate the utility of our new approach to satisfy their needs in domain research. The experts were from two major research areas. One group (GA) had two experts (professors with Ph.D. degrees) working on using machine/deep learning in engineering fields including the study of unmanned vehicles, robots, and their swarms. One professor had developed driving recognition models based on DL and human behavior study. The models were used to build trust between humans and autonomous vehicles. Another group (GU) had three experts (all with Ph.D. degrees and two professors) working in urban study, remote sensing, geography, and land use. They have an interest to understand ADMs and their relations to geographical factors. We conducted individual virtual meetings with each expert for about 1 to 2 hours including system demonstration, test use, and interview for system evaluation. 

First, we introduced the ADMs and video datasets used in the system. We then showed our Web-based VA interface. Next, the experts explored the interface within the NYC area through an internet browser. Finally, they were asked to evaluate the system in several facets by answering questions including:
\begin{itemize}
    \item \textbf{System utility}: \\
    \underline{For GA}: How do you think this system can contribute to the study of autonomous driving models?\\
    \underline{For GU}: How do you think this system can contribute to the study of geographical infrastructure and transportation?
    
    \item \textbf{Comparison:} \\
    How do you compare this system with existing visualization approaches in your field?
    
    \item \textbf{Usefulness of visualizations:}
    How do you evaluate the visualization functions in support of data analysis? 

    \item \textbf{Limitation:}
    What are the limitations of the system and visualization functions?
    
    \item \textbf{Suggestions:}
    How do you suggest us to further improve the system? 
\end{itemize}

\subsection{Evaluation Results}
We collected and summarized experts' answers to the questions.
\begin{itemize}
    \item \textbf{System utility}: \\
    \underline{For GA}: The experts agreed that the system could contribute to the research in ADM. They identified that the system could help researchers analyze performances and identify the causes of unqualified data, or wrong labels, or threshold settings for ADMs. They claimed the system ``... be really helpful to analyze model bad performance, and potential risk ...''. With this system the models ``will be continuously improved for performance and reliability''. They pointed ``One of the biggest challenges in an ADM model is to identify where a model is not performing well. In my opinion, the system will be highly appreciated by the ADM-based researchers'' to analyze if there is a need to collect training datasets at different locations.
    
    \underline{For GU}:  The experts believed that this work has the potential to be used in infrastructure and transportation study. They had not realized that such large-scale open-source datasets (e.g., BDD) were utilized by geographical and urban researchers. They said this system may help domain users utilize such spatial and video datasets in a new way. Furthermore, they liked the system since it could ``downscale spatial analytic to fine-level and space-time resolution within a highly visual context''. ``This is more understandable to domain researchers on urban infrastructure and transportation.''

    \item \textbf{Comparison:} \\
    \underline{For GA}: The experts said there were no existing visualization systems that did the same work. One professor claimed that ``I think this is a cutting edge ADM system''. 
    
    \underline{For GU}: The experts did not realize any visualization system using ADMs and spatial video datasets for geography-based study.
    \\
    All groups liked the new approach of this system and thought it could contribute to their fields with the new datasets and models. One expert noticed that ``the system interactively visualizes the algorithms in the urban scene ...''.
    
    \item \textbf{Usefulness of visualizations:}
    
    All the experts in both groups agreed that the system was very useful in ADM data analysis with the visualization functions. 
    They preferred to use spatial condition filter and model metric filters. The mostly liked functions were the ability that they could study ADMs and locations with respect to environmental attributes such as day, night, rainy, or snowy.

    \underline{For GA}: The experts found it was very useful to observe the errors of different models in the timeline view, which provides a very detailed analysis. GA experts also liked region-based study: ``I like the feature as it allows us to explore the performance of an ADM for any specific region of a city.''
    
    \underline{For GU}: The experts agreed that the system allowed them to work on ADM and spatial video data, with familiar functions as in many GIS tools. They mentioned the algorithm performance analysis could help urban scientists understand the impacts of urban environment settings on driving behaviors. ``The system is also useful to identify location feature with clustered incidents in very few locations.''
    
    \item \textbf{Limitation:}
    
  \underline{For GA}: The experts would like the system to integrate object detection or other computer vision-based techniques (e.g. semantic segmentation) for further analysis of video contents. For example, moving humans (e.g. pedestrians) in the scene played critical roles in driving decision making, which could be used in the system. They also realized domain users might need some efforts to learn and understand the visualization system. They also suggested extending the system to more open datasets and cities.
  
  \underline{For GU}: The experts would like the visualization system to include available road conditions, traffic congestion, the dilapidation of houses in the neighborhood, etc. One expert also pointed that the system did not allow users to upload their own data.
    
    \item \textbf{Suggestions:}
    
    The experts suggested us to address the above limitations. Moreover, some of them suggested that ``interpretation of unsatisfied performance such as cause or consequence generation could be improved.'' One expert mentioned that it would be helpful if researchers were allowed to make changes in an existing model and see performance changes.  An GU expert also suggested adding a data fusion function to integrate various urban datasets to enrich the urban background information for the ADM comparison. Some also suggested a set of visualization details such as adding more labels and explanations to the terms and functions. 

\end{itemize}

\section{Discussion and Conclusion}
In this paper, we develop and evaluate new visualization techniques for analyzing ADM prediction data within a geographical environment. It presents a new VA system which can be used by domain experts to enhance their data analytic capability of large-scale ADM data. The system can be further improved in several directions. 

\noindent \textbf{System Scalability:} 
The system is designed for the BDD dataset with some limitations of scalability. First, we currently study three ADMs. Although we expect a few other models can be added directly, if more than 10 models are to be studied together, the system interface should be re-designed to accommodate them. Second, the driving actions are limited to four. If more actions are interested such as ``slightly turn left or right’’, the system should easily handle it. However, if continuous action predictions of wheel angles and speed ranges need to be analyzed, the system will need to be revised. Third, the system cannot integrate more AD attributes such as multiple camera inputs and IMU sensor data, which would be our future study topics.

When the size of models, actions, and data increase, it may become overwhelming for users to conduct effective studies. Excessive exploration efforts may impair their capability to form insights. Therefore, automatic recommendation algorithms and visualizations may need to be developed for the system. 

\noindent \textbf{ADM refinement:} The system mostly focuses on data analytics of ADM predictions with geo-visualization tools. However, it does not solve the problem of how the spatial context information can be used to improve the perception models with actionable insights. Moreover, the system is not designed for on-the-fly model debugging and refinement. In the future, VA tools can be developed to integrate spatial information with the design of such neural networks by AD experts.

\noindent \textbf{Video content:} Video contents are used in this work as image frames coupled with predictions and locations. The spatial videos can be further processed by computer vision tools, such as DL-based segmentation and object detection. This information can be further integrated into the VA system. For example, pedestrians may be extracted and add a new aspect in visual analysis to better explain driving predictions.

\noindent \textbf{Data integration:} There exist many other types of urban datasets which may be incorporated into the ADM analytical system. In the current system, the spatial conditions include street types, weather, and video recording time. We could further include traffic data, neighborhood data (e.g., highrise buildings), and other influential information in the analysis. 

\noindent \textbf{Domain research:} The emerging ADM datasets including large-scale spatial video data may become a useful and interesting resource for research of a variety of urban scientists. As we can foresee the prevalence of autonomous driving vehicles in a near future, urban researchers either from transportation studies, urban planning, or social sciences want to know more about ADMs and their influences. Our system may give them a visualization tool to start the work, by integrating their domain-specific problems.

The real world AD system can be much more complex with different perception models with multiple sensor inputs. In future work, we first plan to address the limitations and suggestions pointed out by the domain experts in evaluation. We will also extend the use of this tool to more ADM scenarios. We will further make new technical development as discussed in this section.

\acknowledgments{The authors wish to thank the anonymous reviewers. 
This work was partly supported by NSF-1739491 and Kent State University graduate assistantship. } 

\bibliographystyle{abbrv}  

\bibliography{Driving,LRP}
\end{document}